%% file: paper.tex
\documentclass[sigplan,screen,nonacm]{acmart}
\makeatletter
\@printpermissionfalse
\makeatother
\renewcommand\footnotetextcopyrightpermission[1]{}
\usepackage[normalem]{ulem}

\usepackage{booktabs}   
\usepackage{enumitem}   
\usepackage{threeparttable}  
                        
\usepackage{listings}
\usepackage{algorithm}  
\usepackage{algorithmicx}
\usepackage{algpseudocode}  
\usepackage{subfig}
\usepackage[inkscapelatex=false]{svg}

\usepackage{multirow}
\usepackage{tikz}
\usepackage[hyphenbreaks]{breakurl}
\usepackage{soul}
\usepackage{pifont}

\usepackage{bbding}
\usepackage{makecell}

\lstdefinelanguage{java}{
  morekeywords={abstract,case,catch,class,def,%
    do,else,extends,false,final,finally,%
    for,if,implicit,import,match,mixin,%
    new,null,object,override,package, public,%
    private,protected,requires,return,sealed,%
    super,this,throw,trait,true,try,%
    type,int,double,while,pnew, persistable, atomic, durable\_root},
  otherkeywords={=>,<-,<\%,<:,>:,\#,@},
  sensitive=true,
  morecomment=[l]{//},
  morecomment=[n]{/*}{*/},
  morestring=[b]",
  morestring=[b]',
  morestring=[b]""",
}

\lstnewenvironment{q1}[1][]
{\lstset{
frame=none,
language=java,
aboveskip=3mm,
belowskip=3mm,
showstringspaces=false,
columns=flexible,
breaklines=true,
basicstyle={\footnotesize\ttfamily}, 
tabsize=4,  
numbers=left,
xleftmargin=2em,
framexleftmargin=2em,
#1} }
 {}
 
\begin{document}

\newif\ifarxivrename
\arxivrenametrue

\newcommand{\submissiontitle}{RunC or RunV? No! Paravirtualized Containers Go Beyond the Isolation--Performance Trade-off}
\newcommand{\arxivtitle}{ParaCell: Paravirtualized Secure Containers with Lightweight Intra-Container Isolation and Intent-Driven Memory Management}
\newcommand{\submissionsys}{RunPV}
\newcommand{\arxivsys}{ParaCell}

\newcommand{\papertitle}{%
  \ifarxivrename
    \arxivtitle
  \else
    \submissiontitle
  \fi
}

\title{\papertitle}

\renewcommand{\shortauthors}{Wu et al.}


\author{Yiyang Wu}      
\authornote{Contributed equally to this work.}
\affiliation{
 \department{Institute of Parallel and Distributed Systems}              
 \institution{Shanghai Jiao Tong University}            
 \city{Shanghai}
 \country{China}
}
\email{adminstrator@sjtu.edu.cn}       

\author{Xunjie Wang}
\authornotemark[1]
\affiliation{
 \department{Institute of Parallel and Distributed Systems}              
 \institution{Shanghai Jiao Tong University}            
 \city{Shanghai}
 \country{China}
}
\email{vectorxj@sjtu.edu.cn}       

\author{Jinyu Gu}
\affiliation{
 \department{Institute of Parallel and Distributed Systems}              
 \institution{Shanghai Jiao Tong University}            
 \city{Shanghai}
 \country{China}
}
\email{gujinyu@sjtu.edu.cn}

\author{Haibo Chen}
\affiliation{
 \department{Institute of Parallel and Distributed Systems}              
 \institution{Shanghai Jiao Tong University}            
 \city{Shanghai}
 \country{China}
}
\email{haibochen@sjtu.edu.cn}       

\newcommand{\runc}{RunC}
\newcommand{\runv}{RunV}
\newcommand{\sys}{%
  \ifarxivrename
    \arxivsys
  \else
    \submissionsys
  \fi
}
\newcommand{\us}{\textmu s}

\newcommand*\circled[1]{\tikz[baseline=(char.base)]{
            \node[shape=circle,draw,inner sep=0.1pt, minimum size=9pt] (char) {\footnotesize #1};}}

\newenvironment{myitemize}
  {\begin{list}{\labelitemi}{\itemsep1pt \topsep2pt \parsep0.00in
  \partopsep=0pt \leftmargin1.4em}}%
  {\end{list}}

\newcommand{\ra}[1]{\renewcommand{\arraystretch}{#1}}
\newcommand{\stitle}[1]{\vspace{0.5ex}\noindent{\bf #1}}

\newcommand{\pie}[1]{%
\begin{tikzpicture}
 \draw (0,0) circle (1ex);\fill[rotate=90] (1ex,0) arc (0:#1:1ex) -- (0,0) -- cycle;
\end{tikzpicture}%
}

\newcommand{\TODO}[1]{\textcolor{red}{[TODO: #1]}}
\newcommand{\wyy}[1]{\textcolor{blue}{[WYY: #1]}}
\newcommand{\gates}{XGates} 
\newcommand{\pager}{Pager}

\date{}
\input{abs}




\maketitle


\input{intro}
\input{motivation}

\input{overview}

\input{design}

\input{impl}
\input{eval}
\input{related-work}
\input{conclusion}



\bibliography{paper}




\input{appendix}

\end{document}

%% file: abs.tex
\begin{abstract}
Secure containers isolate each container with its own kernel, mitigating shared-kernel attacks prevalent in traditional container systems.
However, existing designs still face a fundamental isolation--performance trade-off.
Nested-cloud deployments amplify the cost of VM exits and page-table management,
while emerging agentic workloads expose bursty memory demand that requires fine-grained elasticity.
We attribute this trade-off to two root causes.
First, existing designs lack lightweight intra-container isolation primitives for frequent container user--kernel transitions.
Second, the host treats container memory management as opaque,
forcing reactive secondary faults and coarse-grained huge page mappings to amortize their cost.

This paper presents \sys{}, a paravirtualized secure container runtime built on two insights.
First, intra-address-space hardware protection primitives can provide lightweight intra-container isolation.
\sys{} uses MPK-based \gates{} to isolate the container user and container kernel within a single address space,
turning frequent user--kernel transitions into direct domain switches.
Second, container kernel allocators already encode memory-management intent.
\sys{} introduces \pager{} to interpose on allocation and free events,
batch proactive GPA$\rightarrow$HPA bindings and unbindings,
and avoid reactive shadow page-table faults while preserving fine-grained memory elasticity.

\sys{} is implemented as a drop-in replacement for \runv{}.
Our experiments demonstrate that, across traditional cloud and emerging agent applications, \sys{} reduces latency by up to 57\% and 79\% over PVM, and by up to 33\% and 88\% over \runv{},
in bare-metal and nested setups, respectively.
On agent workloads, \sys{} saves up to 35.6\% memory compared with the state-of-the-art VM memory reclamation technique, HyperAlloc.
\end{abstract}

%% file: intro.tex
\section{Introduction}
\label{sec:intro}

Containers are the dominant deployment unit in today's cloud.
OS-level containers such as \runc{}~\cite{runc} share the host kernel,
delivering near-native performance~\cite{oakes2018sock,wei2023mitosis,huang2024trenv}
but inheriting a broad kernel attack surface~\cite{haq2024sok,jarkas2025container,he2023ebpf}
and cross-container interference~\cite{wang2023characterizing,liu2023kit,volpert2025detecting}.
Secure containers address these limitations by moving from a \emph{shared-kernel} architecture
to a \emph{separate-kernel} architecture: each container runs with its own guest kernel,
so untrusted workloads no longer directly share the host kernel as their primary execution substrate.
This model has become an important foundation for secure multi-tenancy in production clouds,
including systems from AWS~\cite{agache2020firecracker}, Ant Group~\cite{chai2026skernel},
and Microsoft Azure~\cite{alquraan2026drops}.
Kata-style \runv{} containers~\cite{kata,runv} are a representative realization of this model,
using hardware virtualization to enforce the container boundary while preserving Linux compatibility.

Existing secure-container designs, however, face growing pressure in nested-cloud deployments and emerging agentic workloads.
Nested-virtualization deployments, where tenants run their own hypervisors inside leased VMs, make hardware-assisted \runv{} costly:
VM exits and nested EPT faults must be mediated through the outer hypervisor,
substantially increasing the latency of operations on the container boundary~\cite{zur2025accelerating,shi2025hardware,huang2023pvm}.
At the same time, agentic workloads such as OpenClaw~\cite{openclaw} and Hermes~\cite{hermes} are moving to cloud-hosted deployments for easier use and management~\cite{alibaba-openclaw-deploy,tencent-openclaw-deploy,tencent-hermes-deploy}.
Such workloads make sandboxing a first-class requirement because they execute dynamic, partially untrusted tool actions.
Meanwhile, their memory demand is also highly bursty and fluctuating:
prior work reports peak-to-average spikes of up to 15.4$\times$~\cite{zheng2026agentcgroup},
and our traces from SWE-bench runs show a mean 23.3\% waste when memory is accounted at 2MB granularity instead of 4KB granularity
(Figure~\ref{fig:motivation:swe_bench_fragmentation}).
These workloads need secure containers that provide both efficient frequent boundary crossings and fine-grained memory elasticity.

Following this analysis, we attribute the mismatch to two root causes.
The first root cause is the lack of lightweight intra-container isolation primitives.
A secure container must isolate the container user from its container kernel,
while still making common user--kernel transitions such as syscalls and page faults cheap.
Hardware-virtualized designs~(e.g., \runv{}) preserve efficient in-guest transitions,
but bind this intra-container boundary to hardware virtualization,
which becomes costly and inflexible under nesting.
Software-virtualized designs~(e.g., PVM~\cite{huang2023pvm}) avoid that dependence,
but isolate the container user and container kernel with separate address spaces,
placing page-table switches on frequent transition paths.

The second root cause is the opacity of container memory-management intent to the host.
Traditional virtualization exposes the guest's page-table state rather than its allocator intent.
The host therefore discovers actual memory demand reactively, through secondary faults such as EPT violations or shadow faults.
This, however, creates a granularity dilemma.
Fine-grained 4KB backing preserves good memory utilization but turns page faults into frequent host--guest transitions,
whereas huge-page backing amortizes those faults but wastes memory for bursty and sparse workloads.

We therefore propose \sys{}, a paravirtualized secure-container runtime that addresses these two root causes through two insights.
The first insight is that recent intra-address-space protection primitives can provide the missing lightweight intra-container isolation layer.
\sys{} therefore uses MPK-based \gates{} to place the container user and container kernel in one address space while separating them into MPK-protected domains.
This changes frequent user--kernel transitions from address-space switches into lightweight domain switches,
preserving the isolation abstraction without relying on hardware virtualization support.

The second insight is that the container kernel already knows the memory-management intent that the host lacks.
Kernel allocators define the lifecycle of physical pages:
allocation means a page is about to become active,
and free means the backing memory can be reclaimed.
\sys{} introduces \pager{} to expose these allocator events as an explicit contract between the container kernel and the host.
Rather than waiting for secondary faults to infer demand,
\pager{} proactively binds and unbinds container pages to host pages.
It directly maintains host-side mappings from allocator intent,
rather than synchronizing them reactively through shadow page-table faults,
preserving efficient management while keeping host allocation close to actual guest use at 4KB granularity.

We implement \sys{} as a drop-in replacement for \runv{} in the containerd ecosystem and evaluate
traditional cloud applications (both nested and non-nested deployments) and emerging agent workloads (SWE-bench~\cite{jimenez2023swe} with Claude Code, SkillsBench~\cite{li2026skillsbench}).
Across cloud and agent applications, \sys{} reduces latency by up to 57\% and 79\% over PVM,
and by up to 33\% and 88\% over \runv{}, in bare-metal and nested setups, respectively.
On agent workloads, \sys{} saves up to 35.6\% memory over the state-of-the-art reclamation technique HyperAlloc~\cite{wrenger2025hyperalloc},
with negligible memory waste of less than 0.4\% on average.
In summary, this paper contributes:
\begin{itemize}[leftmargin=*,nosep]
\item An analysis of nested-cloud and agent workloads that identifies two structural causes of secure-container overhead:
  heavyweight intra-container isolation and opaque guest memory-management intent
  (\S\ref{sec:motivation}; \S\ref{sec:overview:analysis_of_isolation_performance_tradeoffs}).
\item The design and implementation of \sys{}, which combines MPK-based \gates{} for lightweight guest user--kernel transitions
  with \pager{} for intent-driven, fine-grained container memory management
  (\S\ref{sec:overview}--\S\ref{sec:impl}).
\item An evaluation showing that \sys{} improves secure-container performance under both bare-metal and nested deployments
  while providing fine-grained memory elasticity
  (\S\ref{sec:eval}).
\end{itemize}

%% file: motivation.tex
\section{Background and Motivation}
\label{sec:motivation}

\subsection{Secure Containers}
\label{sec:motivation:sec_ctnr}

Conventional OS-level containers such as \runc{}~\cite{runc} rely on a \emph{shared kernel} architecture,
achieving low overhead at the cost of fundamental isolation and scalability limitations.
First, the large kernel attack surface shared by all tenants has long been recognized as a major source of container escapes and sandbox breaks~\cite{tunde2019study,manco2017my,liu2023kit,shi2025hardware}.
Second, co-located containers interfere through shared kernel code paths and global kernel state,
leading to cross-tenant performance interference~\cite{swift2025locked},
especially on modern many-core servers~\cite{boyd2010analysis,hoffmann2011everything,xiao2025exploiting}.

Secure containers avoid these issues by moving to a \emph{separate kernel} architecture that provides each container with its own kernel.
For this reason, major cloud vendors such as AWS~\cite{agache2020firecracker}, Ant Group~\cite{chai2026skernel},
and Microsoft Azure~\cite{alquraan2026drops} have adopted secure containers as an important deployment model in cloud environments.
In this paper, we therefore restrict our scope to \emph{secure containers}.

\subsection{Why \runv{} Falls Short}
\label{sec:motivation:why_runv}

Hardware-assisted virtualization-based secure containers (\runv{}~\cite{runv}), such as Kata Containers~\cite{kata},
address the isolation limitations of shared-kernel containers by placing each container inside a VM with its own kernel,
using hardware virtualization support such as Intel VT-x~\cite{uhlig2005intel} and AMD-V~\cite{virtualization2005secure}.
However, this design increasingly falls short in important scenarios such as nested clouds and agentic workloads.

\subsubsection{Nested Clouds}
\label{sec:motivation:nested}

First, nested cloud settings dramatically amplify the cost of hardware virtualization,
making \runv{} increasingly unattractive for modern IaaS deployments~\cite{shi2025hardware,huang2023pvm}.
In this setting, secure containers (L2 VMs) run inside tenant VMs (L1) leased from an IaaS provider (L0),
a configuration commonly used to obtain stronger isolation and Kubernetes-compatible cluster management~\cite{huang2023pvm}.
However, \runv{} is inefficient when it cannot directly access hardware virtualization support, as in nested clouds.

\stitle{Nesting amplifies hardware-assisted virtualization overheads.}
\runv{} relies on hardware virtualization support such as extended page table (EPT) and VM control structure (VMCS) for efficient memory isolation and CPU virtualization.
Under nesting, however, these mechanisms are only directly available to the outermost hypervisor.
As a result, the L0 hypervisor must virtualize the virtualization interface for the L1 guest,
so privileged operations from L2 containers can no longer be handled by the L1 guest hypervisor alone and must instead be mediated by L0.

This extra indirection penalizes two critical paths.
First, each VM exit from an L2 container must be forwarded through L0 to L1 and then resumed along the reverse path,
adding \emph{two} extra world switches to each VM exit round trip~\cite{zur2025accelerating}.
Prior work shows that it can reduce throughput by up to 4.3\texttimes{} for I/O-intensive applications~\cite{shi2025hardware}.
Second, EPT management is even more expensive.
Beyond the same L0-mediated VM-exit path for L2 EPT fault handling,
nested memory virtualization also requires L0 to maintain a shadow EPT (SEPT) for each L2,
further increasing the cost of L2 EPT faults~\cite{zur2025accelerating,huang2023pvm}.
Prior work reports that nested EPT fault handling introduces \emph{four} extra world switches and increases container startup latency by 114\% compared with the bare-metal case~\cite{zur2025accelerating}.

\subsubsection{Agentic Workloads}
\label{sec:motivation:agent}

Second, emerging agentic workloads expose a fundamental mismatch with \runv{}'s memory management model.
Agent systems such as Codex~\cite{openai-codex}, Claude Code~\cite{anthropic-claude-code},
OpenClaw~\cite{openclaw}, and Hermes~\cite{hermes} integrate LLM reasoning with autonomous tool execution:
the agent observes workspace state, asks the LLM to generate tool actions, and executes those actions as subprocesses.
This execution model is now moving from local developer environments into public clouds:
Alibaba Cloud and Tencent Cloud offer OpenClaw and Hermes deployment services~\cite{alibaba-openclaw-deploy,tencent-openclaw-deploy,tencent-hermes-deploy},
where tenants rent isolated environments to run agents.
Since these generated actions are dynamic, partially untrusted, and externally sourced,
agent execution requires sandboxing as an essential property~\cite{openclaw-sandbox,zheng2026agentcgroup}.
However, their memory characteristics make \runv{} ill-suited as the sandbox substrate.

\begin{figure}[htb]
    \centering
    \begin{minipage}[t]{0.46\columnwidth}
        \centering
        \includegraphics[width=\linewidth]{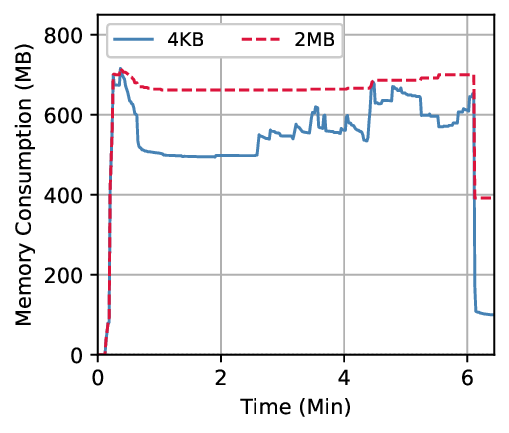}

        (a)
        \label{fig:motivation:swe_bench_fragmentation_trace}
    \end{minipage}
    \hfill
    \begin{minipage}[t]{0.50\columnwidth}
        \centering
        \includegraphics[width=\linewidth]{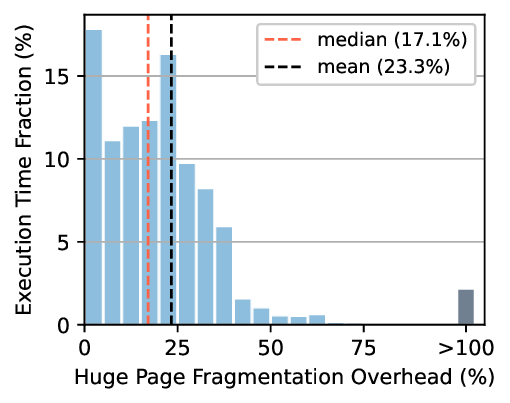}

        (b)
        \label{fig:motivation:swe_bench_fragmentation_distribution}
    \end{minipage}
    \caption{
        Memory characteristics of SWE-bench agent execution.
        (a) In-use guest memory at different page granularities over execution time.
        (b) Distribution of execution time fraction across memory waste levels.
        }
    \label{fig:motivation:swe_bench_fragmentation}
\end{figure}

\stitle{Bursty and fluctuating agentic workloads demand fine-grained memory elasticity.}
We profile agent memory characteristics by sampling at 1Hz during SWE-bench~\cite{jimenez2023swe} runs with Claude Code.
Figure~\ref{fig:motivation:swe_bench_fragmentation}(a) shows the memory-usage time series for a sampled run,
illustrating that agent memory demand is highly bursty and fluctuating.
Prior work reports peak-to-average spikes of up to $15.4\times$ within 1--2 seconds~\cite{zheng2026agentcgroup},
and we also observe similar bursty spikes during agent execution.
Furthermore, Figure~\ref{fig:motivation:swe_bench_fragmentation}(b) shows that, across all sampled SWE-bench executions,
memory waste, defined as the extra memory consumed by 2MB pages relative to 4KB pages, has a mean of 23.3\% and a median of 17.1\%.
Consequently, these patterns demand two properties from the container runtime:
(1) the host must allocate and reclaim container memory quickly (\emph{high resizing performance}); and
(2) the gap between guest-used and host-allocated memory must remain small (\emph{high memory utilization}).

\runv{} instead faces a \emph{utilization--performance trade-off}.
It typically relies on huge EPT pages for performance,
which can reduce page walk latency by 40\% and improve performance of typical memory-intensive applications by 12\%~\cite{li2023fhpm}.
However, it makes memory resizing coarse-grained and increases fragmentation,
since partially used huge pages cannot be reclaimed efficiently~\cite{wrenger2025hyperalloc}.
Consequently, \runv{} struggles to simultaneously provide both high resizing performance and memory utilization for agent workloads.

\subsection{Existing Secure-Container Studies}
\label{sec:motivation:existing}

We study existing secure container designs along three dimensions: performance, memory elasticity, and compatibility.
For performance, we focus on frequent cross-boundary operations such as syscalls and page faults.
For memory elasticity, we examine whether a design can provide fast resizing while preserving high memory utilization.
For compatibility, we consider whether it requires software or hardware modifications.
Table~\ref{tab:motivation:existing} summarizes the comparison.

\begin{table}[htbp]
  \centering
  \caption{Comparison of existing secure container designs. BM: bare-metal. NST: nested.}
  \footnotesize
  \label{tab:motivation:existing}
  \ra{1.2}
  \setlength{\tabcolsep}{4pt}
  \begin{threeparttable}
  \begin{tabular}{l|cc|c}
    \hline
    & \multicolumn{2}{c|}{\textbf{Performance}}
    & \multirow{2}{*}{\makecell{\textbf{Memory}\\\textbf{Elasticity}}} \\
    \cline{2-3}
    \textbf{System}
    & \textbf{Syscall} & \textbf{Page~Fault}
    & \\
    \hline
    \runv{}~\cite{kata}                      & \mbox{\pie{360}}                      & \mbox{\pie{180}~(BM)~\pie{90}~(NST)} & \mbox{\pie{180}} \\
    PVM~\cite{huang2023pvm}                  & \mbox{\pie{270}}                      & \mbox{\pie{180}}                     & \mbox{\pie{180}} \\
    CKI~\cite{shi2025hardware}               & \mbox{\pie{360}}                      & \mbox{\pie{360}}                     & \mbox{\pie{0}}   \\
    X-Container~\cite{shen2019x}             & \mbox{\pie{360}}                      & \mbox{\pie{180}}                     & \mbox{\pie{0}}   \\
    gVisor~\cite{gvisor}                     & \mbox{\pie{360}~(BM)~\pie{180}~(NST)} & \mbox{\pie{180}}                     & \mbox{\pie{360}} \\
    LiteShield~\cite{manakkal2025liteshield} & \mbox{\pie{270}}                      & \mbox{\pie{360}}                     & \mbox{\pie{360}} \\
    \hline
    \textbf{\sys{}} (ours)                   & \mbox{\pie{360}}                      & \mbox{\pie{360}}                     & \mbox{\pie{360}} \\
    \hline
  \end{tabular}
  \end{threeparttable}
\end{table}

\stitle{PVM.}
PVM~\cite{huang2023pvm} eliminates the hardware virtualization dependency by adopting a paravirtualization architecture:
the host resides in kernel mode while both guest user and guest kernel execute in user mode,
each with separate page tables.
As a result, each syscall and page-fault forwarding path must perform two and three page-table switches, respectively.

\begin{figure}[htb]
    \centering
    \includegraphics[width=0.8\columnwidth]{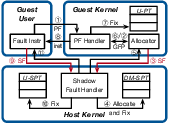}
    \caption{
        Additional shadow faults incurred by shadow paging during anonymous page fault handling.
        Page fault forwarding~(\protect\circled{1}) and emulation of guest page table writes~(\protect\circled{7}) are omitted for brevity.
        PF: page fault. SF: shadow fault.
        GFP: \texttt{get\_free\_pages}. U-PT: page table of the faulting process.
        U-SPT: shadow page table of the faulting process. DM-SPT: shadow page table of kernel direct mapping.
    }
    \label{fig:motivation:sec_fault}
\end{figure}

For memory virtualization, PVM relies on KVM shadow paging,
which maintains a shadow page table (SPT) for each guest page table (GPT) and synchronizes the GPT with the SPT during shadow faults.
Figure~\ref{fig:motivation:sec_fault} illustrates the workflow for handling an anonymous page fault from the guest user.
Specifically, it first traps into the guest kernel's page fault handler~(\circled{1}).
When the guest kernel allocates and clears a page~(\circled{2}) through the kernel direct mapping~(a.k.a.\ linear mapping),
the corresponding guest page has not yet been backed by an HPA.
This triggers a shadow fault~(\circled{3}) to synchronize the direct mapping GPT with its SPT~(DM-SPT).
The host kernel then allocates a host page for the GPA and installs it into the DM-SPT~(\circled{4}).
However, this only fixes the GVA of the direct mapping (i.e., fixing only the DM-SPT).
Then the page is mapped to the faulting user GVA~(\circled{7}) in U-PT,
after which the guest kernel invokes \texttt{iret} to resume user execution at the same faulting instruction.
However, this access triggers another shadow fault~(\circled{9}) due to missing U-SPT entry,
causing an additional GPT-to-SPT synchronization overhead for each new GVA alias of the same GPA.
Together, PVM's complex page fault handling causes up to a 2.1\texttimes{} slowdown compared with \runc{} on memory-intensive applications in our evaluation~(\S\ref{sec:eval:memory-intensive}).

PVM can leverage the conventional virtio-balloon driver~\cite{waldspurger2002memory,schopp2006resizing} to return memory to the host,
but it is still subject to the utilization--performance trade-off (\S~\ref{sec:motivation:agent}).

\stitle{CKI.}
CKI~\cite{shi2025hardware} extends the Protection Keys for Supervisor (PKS)~\cite{mpk} hardware feature to create an isolated intra-kernel domain without relying on hardware virtualization.
It runs the guest kernel as a sandboxed module in kernel mode, thereby eliminating the page-table switches between guest user and guest kernel that PVM incurs.
Additionally, CKI preallocates memory in fixed segments to remove the two-stage address translation entirely.
Together, these techniques improve syscall and page fault performance over traditional software virtualization.
However, this design is not deployable on commodity processors because it requires hardware modifications to PKS,
and its fixed-segment preallocation also prevents fine-grained memory reclamation.

\stitle{X-Container.}
X-Container~\cite{shen2019x} adopts a Xen-PV-based LibOS architecture,
in which the guest kernel and user application are colocated without isolation.
By using dynamic binary rewriting to transform system calls into direct function calls,
it achieves low syscall overhead.
However, its threat model is incompatible with agentic applications.
For example, agents increasingly operate browser environments~\cite{openai-computer-use,claude-code-computer-use,playwright-mcp},
while modern browsers rely on processes to isolate mutually untrusted web content~\cite{chromium-site-isolation,firefox-site-isolation}
and provide crash isolation across browser components~\cite{chromium-multi-process}.
Additionally, page table operations on page faults still require hypercalls to the hypervisor.
Moreover, it statically allocates all guest memory to avoid shadow paging overhead,
precluding memory elasticity.


\stitle{gVisor.}
gVisor~\cite{gvisor} adopts an application kernel, called the Sentry, which implements a Linux-like interface for each sandboxed container.
By moving the kernel interfaces normally provided by the host kernel into a per-container Sentry,
gVisor reduces the host kernel attack surface while retaining a process-like resource model.
In practice, gVisor is commonly deployed with two platforms.
When hardware-assisted virtualization is not available (e.g., in nested clouds),
gVisor places the Sentry in userspace and leverages Systrap~\cite{gvisor-systrap} to redirect syscalls and page faults to the Sentry, incurring additional IPC overhead.
Otherwise, gVisor uses KVM~\cite{kivity2007kvm} so that the Sentry acts as both the guest OS kernel and the VMM,
reducing costly guest user/kernel switches (e.g., syscall transitions).
However, page faults still involve additional EPT violations, as in \runv{}.
For memory elasticity, the Sentry can use \texttt{fallocate} to directly notify the host about freed memory, avoiding the overhead of memory ballooning~\cite{waldspurger2002memory,schopp2006resizing}.
However, because gVisor reimplements Linux kernel interfaces in the Sentry, it may lack the full compatibility and kernel-level optimizations of native Linux.

\stitle{LiteShield.}
LiteShield~\cite{manakkal2025liteshield} adopts a microkernel architecture that delegates most application syscalls to user-space system services via shared-memory IPC.
This design provides low syscall latency by avoiding context switches, but relies on polling service threads and thus hurts CPU utilization.
Its page faults and memory elasticity remain efficient because memory management is still handled by the host kernel.
However, compatibility is limited by the need to reimplement or port existing stacks into the user-space service layer.

%% file: overview.tex
\section{Overview}
\label{sec:overview}

\subsection{Analysis of Isolation--Performance Trade-offs}
\label{sec:overview:analysis_of_isolation_performance_tradeoffs}
Important cloud workloads~(\S\ref{sec:motivation:nested}, \S\ref{sec:motivation:agent}) place simultaneous isolation and performance demands on secure containers along two axes:
(1) they require intra-container isolation while preserving efficient cross-boundary performance, and
(2) they require fine-grained container memory isolation while delivering efficient memory resizing performance.
We refer to these tensions as the \emph{isolation--performance trade-offs} and attribute them to two root causes.

\stitle{Root cause \#1: The lack of lightweight intra-container isolation primitives.}
A secure container requires two properties within the container:
isolation between the container user and its kernel, and efficient transitions across this user--kernel boundary.
The former preserves the conventional privilege separation expected by applications,
while the latter keeps frequent cross-boundary operations~(e.g., syscalls and page faults) efficient.
Existing VM-based secure containers choose between two imperfect isolation mechanisms.
Hardware virtualization preserves efficient in-guest user--kernel transitions,
but binds the isolation to hardware support and is inflexible in environments such as nested clouds~(\S\ref{sec:motivation:nested}).
Software virtualization avoids this rigid dependence,
but makes frequent user--kernel transitions inefficient.

\stitle{Root cause \#2: The opacity of guest memory management intent.}
Traditional virtualization virtualizes the MMU and page table abstraction.
However, this abstraction does not reflect actual memory usage:
a guest kernel typically establishes a direct mapping~(a.k.a.\ linear mapping) that covers its entire GPA range at boot time,
so all guest page table entries are populated long before any page is actually used by the guest kernel or user process.
The host, treating the guest as an opaque VM, cannot distinguish a pre-mapped but idle page from one actively in use from page-table operation interfaces.
Therefore, the host relies on a reactive mechanism, on-demand paging via \emph{secondary faults}~(e.g., EPT violations and shadow faults),
to discover actual access patterns.
This creates a fundamental granularity trade-off in host-side memory management:
fine-grained backing preserves memory utilization but incurs per-page secondary fault overhead,
while huge-page backing amortizes this overhead but exacerbates memory fragmentation.

\subsection{Architecture Overview}
\label{sec:overview:architecture}

\begin{figure}[htbp]
    \centering
    \includegraphics[width=0.9\columnwidth]{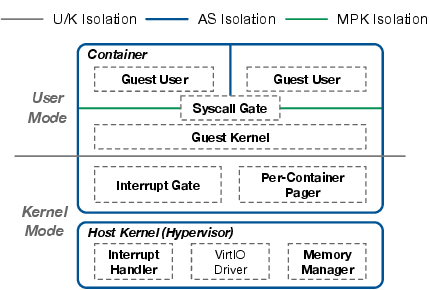}
    \caption{\sys{} architecture. U/K: user/kernel privilege. AS: address space. MPK: Memory Protection Key.}
    \label{fig:overview:runpv}
\end{figure}

We propose \sys{}, a paravirtualized (PV) secure container runtime that addresses the two root causes.
As shown in Figure~\ref{fig:overview:runpv}, \sys{} preserves the VM-like secure container abstraction of a separate guest kernel,
while deprivileging the guest kernel to user mode to avoid rigid dependence on hardware-assisted virtualization,
following prior PV secure containers~\cite{shen2019x,huang2023pvm}.
Within each container,
\sys{} preserves isolation among multiple user applications and between user applications and the guest kernel,
while enabling efficient transitions between guest user and kernel.
As in prior PV secure container designs, the host kernel acts as the hypervisor:
it schedules containers, handles and forwards interrupts for the guest kernel, and processes VirtIO requests.
Unlike prior work, however, \sys{} provides efficient fine-grained memory management for container memory.
We next introduce the two key insights behind \sys{}.

\stitle{Insight \#1: Using emerging lightweight hardware protection primitives to enlighten intra-container isolation.}
To address root cause \#1, we observe that recent hardware extensions that facilitate intra-address-space level isolation gracefully
fits the requirement of intra-container isolation. MPK\cite{mpk} for x86 is a good example that is capable of presenting
an orthogonal supplement to software virtualization to bypass the costly address space switch overhead. 
To utilize these opportunities, We propose \gates{}~(\S\ref{sec:sgate}) in \sys{} to bridge the guest user and kernel
by adopting such intra-address-space isolation under software virtualization. By unifying the guest's address space, we eliminate the strict
overhead of address space switch between guest user and guest kernel, which is a must for traditional software virtualization. We reduce the interfaces between the host and container as much as possible 
and enforce address-space isolation on the container-host boundary to prevent potential jailbreak attack from the container to the host. To further exploit the optimization opportunities,
we design the \gates{} architecture to perform different switch policies for different control flow classes (syscalls, self-generated exceptions, and external interrupts)
respectively and exploit the optimization opportunities on syscall and self-generated exception paths.

\stitle{Insight \#2: Exposing guest kernel memory management intent to the host enables efficient fine-grained container memory management.}
To address root cause \#2, we observe that the guest is not an arbitrary process,
as it runs a kernel whose page allocator already knows which pages will be used and when.
Therefore, \sys{} leverages \pager{}~(\S\ref{sec:mem}) to bridge this semantic gap by interposing at the memory management interface of the guest kernel,
where pages transition between allocation and free through well-defined helper functions.
These lifecycle events are synchronous and intentional,
allowing \pager{} to bind GPA$\rightarrow$HPA mappings at allocation time.
This enables a new form of shadow page table:
instead of a lazily synchronized copy repaired by shadow faults,
\sys{} uses it as the direct target of page-table updates.
When the guest kernel installs a mapping,
\pager{} translates the GPA to its bound HPA and writes the entry directly into the shadow page table.
When pages are freed, \pager{} correspondingly unbinds GPA$\rightarrow$HPA mappings and returns the backing pages to the host.
This design collapses two-stage translation into a single stage to eliminate secondary faults,
and keeps host-allocated memory tightly aligned with guest usage at 4KB granularity.
To further reduce context-switch overhead,
\sys{} places \pager{} in kernel mode to reduce per-switch overhead and batches related hypercalls to amortize VM-exit overhead.

%% file: design.tex
\section{Detailed Design}
\label{sec:design}

\subsection{Lightweight Intra-container Isolation}
\label{sec:sgate}

\begin{figure}[htbp]
    \centering
    \includesvg[width=\columnwidth]{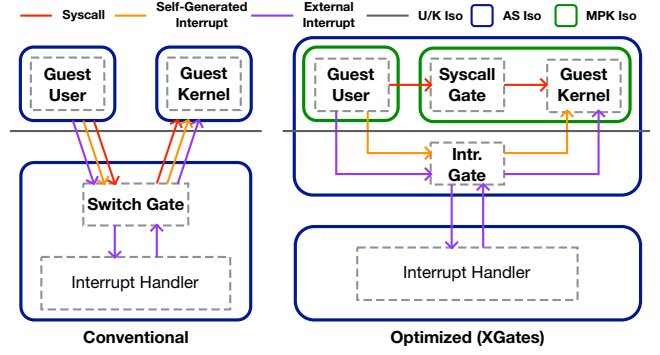}
    \caption{\sys{}'s control-flow-aware domain-switch architecture compared with traditional PV switch gates design. U/K: user/kernel. AS: address space. MPK: Memory Protection Keys. Iso: isolation.}
    \label{fig:design:gates}
\end{figure}

\stitle{Moving towards intra-address-space isolation.}
Section~\ref{sec:motivation} showed that address-space-based guest user--kernel isolation places expensive page-table switches on frequent transition paths.
\sys{} instead co-locates the guest user and guest kernel in one address space and separates their data access with MPK-protected domains.
Figure~\ref{fig:design:gates} illustrates this architecture.
In the Guest User domain (GU), guest-kernel memory is non-readable and non-writable; in the Guest Kernel domain (GK), the kernel has full memory visibility.
The design invariant is that GU--GK transitions occur only through controlled gates, which perform the protection-key update and restore the execution context for the target domain.
With address-space switching removed, frequent user--kernel transitions become direct domain switches instead of privileged page-table switches.

\stitle{Control-flow-aware switch gate design.}
After eliminating address-space switching, the main design question is whether the host must semantically participate in a GU--GK transition.
\sys{} classifies transitions into three cases.
First, syscalls are explicit software transfer points; once syscall sites are rewritten to enter a gate, the host no longer needs to trap on the syscall instruction.
Second, self-generated exceptions such as page faults still arrive through a privileged trap, but the host does not need to emulate their guest-kernel handling, so the trap can be fast-forwarded to GK.
Third, external interrupts require host-side acknowledgment and routing, so they must remain on the conventional host-mediated path.

Based on this classification, each container owns a pair of \gates{} rather than a single unified switch gate.
The Syscall Gate runs in hardware ring-3 and implements the fully deprivileged syscall transition.
Its entry points are introduced by binary rewriting, and its mutable state, including vCPU context pointers, remains in GK-protected memory.
The Interrupt Gate runs in hardware ring-0 and handles privileged event delivery; it fast-forwards self-generated exceptions to the guest kernel, while leaving external interrupts to the host-mediated path.
This split removes host intervention from the hot syscall path, avoids unnecessary host emulation for internal exceptions, and preserves correctness for hardware-driven events.

\stitle{Deprivileging syscalls.}
Turning the syscall path into a deprivileged switch requires replacing the privileged operations used by a conventional \emph{syscall}/\emph{sysret} transition.
Traditional PV designs rely on three such operations:
privileged \emph{cr3} updates switch between isolation domains,
\emph{swapgs}-based bookkeeping locates the vCPU's guest-kernel context (\emph{kernel\_gs}, \emph{kernel\_rsp}, etc.),
and the hardware syscall sequence provides switch atomicity by suppressing interrupt delivery during the transition.

\sys{} replaces these operations with gate-local mechanisms.
MPK domain updates replace \emph{cr3}-based isolation switches.
GK-protected thread-local state replaces \emph{swapgs}-based context lookup, so the gate can always recover the correct vCPU context within the gate if thread--vCPU affinity changes during entries.
Explicit virtual interrupt masking replaces hardware syscall atomicity: the gate blocks virq injection during the critical section by emulating \emph{cli}/\emph{sti} through the interrupt flag in the vCPU context.
Together with syscall-frame manipulation, these mechanisms turn a syscall into a regular gate call that can replace the privileged \emph{syscall} instruction in rewritten binaries.

\begin{figure}[htbp]
    \centering
    \includegraphics[width=0.8\columnwidth]{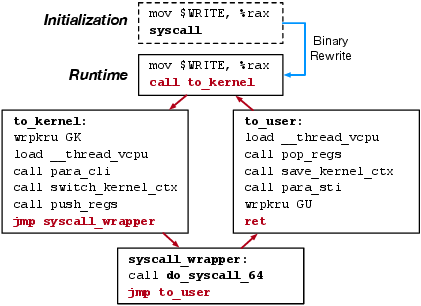}
    \caption{Runtime Syscall Deprivileging inside the Syscall Gate. GK: Guest Kernel. GU: Guest User. \texttt{para\_cli}/\texttt{sti}: emulated \texttt{cli}/\texttt{sti} with vCPU's interrupt flag manipulation.}
    \label{fig:design:syscall_routine}
\end{figure}

During initialization, \sys{} fetches the guest kernel's syscall handler entry, registers the vCPU thread-local mapping,
and exhaustively rewrites syscall invocation sites into hijackable hook points.
Figure~\ref{fig:design:syscall_routine} illustrates the Syscall Gate's runtime behavior.
When an application executes a rewritten syscall site, control is first redirected into the Syscall Gate.

Inside the gate, \texttt{to\_kernel} handles the forward transition. 
It saves user execution context on the stack, switches to the GK memory domain, 
and issues an emulated \emph{cli} to block interrupt delivery in the critical section. 
It then restores guest-kernel context from the registered GK-protected thread-local state
and dispatches to the designated syscall wrapper. When the syscall wrapper returns, control is relayed back to the gate and \texttt{to\_user} performs the reverse transition: 
it saves return state, restores user context, re-enables interrupt delivery, switches the domain key back to GU, 
and resumes execution at the original user return point.
The whole process is self-contained and does not rely on privilege-level switches.

\stitle{Interrupt Fastpath.} 
Unlike the Syscall Gate, interrupt handling still enters the ring-0 Interrupt Gate because interrupt delivery is privilege-mediated. 
At entry, the gate first classifies the interrupt source. For self-generated exceptions (e.g., page faults), \sys{} follows a fast path
that avoids unnecessary host-side handling: it switches to GK, restores the required guest-kernel context, and directly forwards control to the guest interrupt handler. 
For external interrupts, the gate falls back to the conventional host-mediated path, because source acknowledgment and routing remain host responsibilities.
This split preserves correctness for hardware-driven events while minimizing trap-and-switch overhead on internal interrupt paths.

\subsection{Fine-Grained Elastic Memory Management}
\label{sec:mem}

Guest kernel's page allocation events, rather than page table operations, expose the actual memory management intent:
a physical page enters active use only when the guest kernel's allocator hands it out and becomes host-reclaimable only when the allocator takes it back.
These allocation and free events (e.g., Linux kernel primitives \texttt{\_\_get\_free\_pages} and \texttt{free\_pages}) precisely indicate when a GPA requires,
or no longer requires, a backing host physical page.
By exposing these events to the host,
\sys{} can bind a GPA to an HPA at allocation time and unbind it at free time,
replacing reactive, fault-driven two-stage translation with intentional, intent-driven single-stage translation.
Specifically, \sys{} (1) transfers allocation intent through explicit GPA$\rightarrow$HPA (un)binding,
(2) amortizes host page handoff with PCP lists,
and (3) places \pager{} in kernel mode to reduce per-switch overhead,
thereby keeping the page-fault path close to that of a traditional OS.

\begin{figure}[htb]
    \centering
    \includegraphics[width=\columnwidth]{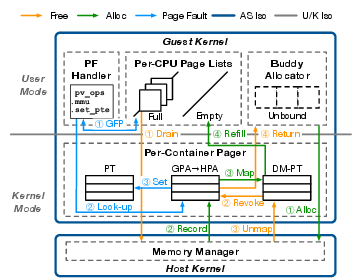}
    \caption{
        \pager{} architecture and workflow. PT: page table. DM-PT: page table managing direct mapping. GFP: \texttt{get\_free\_pages}.
        AS Iso: address space isolation. U/K Iso: user/kernel privilege isolation.
    }
    \label{fig:design:pager}
\end{figure}

\stitle{\pager{} workflow.}
To realize this intent-driven model, \sys{} introduces the \pager{}:
a dedicated memory mediator that translates guest allocation/free operations into GPA$\rightarrow$HPA bindings/unbindings on host pages.
%
To handle the pre-mapped direct mapping,
the guest explicitly registers this region with the \pager{} so that these PTEs are managed according to allocator lifecycle events rather than their static presence.
Figure~\ref{fig:design:pager} illustrates the workflow for page allocation and free, as well as page table updates.

On allocation~(green in the figure), the \pager{} allocates a backing page from the host~(\circled{1}\circled{4}, as explained below),
records the GPA$\rightarrow$HPA binding~(\circled{2}).
It also immediately maps the translated HPA into the direct-mapping PTE~(\circled{3}),
as this allocation typically indicates that the guest kernel will soon access the page through the direct mapping,
for example to clear the page or populate it with file data (e.g., page cache).
On subsequent mappings~(blue) of the same GPA~(e.g., into the page table of a user process),
the guest kernel first calls \texttt{\_\_get\_free\_pages} to obtain a HPA-backed page~(\circled{1}),
after which the \pager{} looks up the corresponding HPA~(\circled{2}) and directly installs it into the shadow page table~(\circled{3}),
eliminating further secondary faults.
When a guest page is freed~(orange), the \pager{} performs the reverse:
it revokes the binding from the table~(\circled{2}), removes the direct mapping~(\circled{3})
and returns the backing host page to the host~(\circled{1}\circled{4}, as explained below).
This keeps host-allocated memory tightly aligned with guest usage at 4KB granularity,
avoiding the inefficiencies of reclamation mechanisms such as memory ballooning,
hotplugging, and delayed host-side reclamation.

\stitle{Efficient mediation.}
Efficient mediation of \pager{} requires reducing two costs: host page handoff between the hypervisor and the \pager{},
and switches to the \pager{} during page-table updates.
In terms of host page handoff,
immediately binding/unbinding on every allocation/free would still incur frequent full context switches to the host.
Hence, \sys{} piggybacks on Linux per-CPU page (PCP) lists,
which serve as a per-core cache between the fast allocation path and the global buddy allocator.
PCP lists allow most allocations and frees to be handled locally without touching the global allocator,
and only refill from or drain back to the buddy allocator in batches.
This design reduces contention on the global allocator and avoids cross-CPU cacheline bouncing.

Specifically, pages in the global buddy allocator remain unbound and have no backing HPA.
Binding and unbinding happen at the boundary between PCP lists and the buddy allocator.
As shown in Figure~\ref{fig:design:pager},
when a PCP list is refilled~(green in the figure) from the buddy allocator,
the \pager{} obtains a batch of host pages from the hypervisor and pre-binds them to the corresponding GPAs;
when pages drain~(orange) from a PCP list back to the buddy allocator,
the \pager{} unbinds the batch and returns the host pages.
Individual allocations~(e.g., \texttt{\_\_get\_free\_pages}) and frees on the hot path merely move pre-bound pages in and out of PCP lists without any host involvement.
This preserves the mature low-contention PCP design,
minimizes intrusive allocator changes, and keeps elasticity off the hot path.
Meanwhile, the memory overhead of pre-bound but not-yet-allocated pages is negligible,
as the size of a single PCP list is typically small, accounting for less than 1\% of the total memory.

After batching host page handoff, \sys{} reduces the remaining switch overhead by choosing where to isolate the \pager{}.
The isolation boundary of the \pager{} is therefore another key design choice.
Isolating it inside the hypervisor would force every mediated operation to cross the expensive guest--host boundary.
Conversely, isolating it within a conventional guest component is insufficient,
as page table updates and TLB maintenance are privileged operations.
\sys{} therefore colocates \pager{} with the guest kernel in the same address space but places it in kernel mode,
separating them only by user/kernel privilege levels.
Privilege-level isolation is sufficient,
as the \pager{} maintains only per-container private state and contains no global data.
By avoiding shared host-kernel mappings,
switching to \pager{} does not require side-channel mitigations such as KPTI and IBRS.

\stitle{Differences from traditional shadow paging.}
With these mechanisms, a page fault~(blue in the figure) remains close to the traditional OS handling path:
the guest kernel allocates a physical page, updates the faulting process's page table, and resumes execution.
Along this path, \pager{} adds only amortized binding overhead and minimal redirection overhead for privileged page-table updates and TLB maintenance.
Unlike traditional shadow paging, \sys{} does not trap-and-emulate guest page table writes or maintain an asynchronous shadow page table,
thereby avoiding the cost of shadow faults.

\stitle{Differences from existing reclamation techniques.}
\pager{} differs from previous work in two ways.
First, it uses guest free events as reclamation signals,
immediately reflecting the guest kernel's memory management intent.
This fine timing suits bursty agentic workloads,
whose memory demand changes quickly across short execution phases~(\S\ref{sec:motivation:agent}).
Second, metadata-driven techniques~(e.g., HyperAlloc~\cite{wrenger2025hyperalloc} and ballooning) scan metadata to detect free pages,
tying reclamation performance to metadata granularity and often requiring coarse amortization.
In contrast, \pager{} avoids metadata scanning entirely while preserving fine granularity.

%% file: impl.tex
\section{Implementation}
\label{sec:impl}

We build \sys{} on PVM~\cite{huang2023pvm}.
On the host-kernel side, we add 3.9K lines of code (LoC) to support the \pager{} and the Interrupt Gate in \gates{}.
On the guest-kernel side, we intrusively modify less than 50 LoC to hook interactions between the buddy allocator and PCP lists,
and add 700 LoC in a separate file to interact with the \pager{} and support the interfaces needed for Syscall Gate initialization.
In terms of \gates{}'s Syscall Gate, we build a dynamic library with about 200 LoC on top of current syscall hook technique~\cite{yasukata2023zpoline}
to implement Syscall Gate initialization and runtime logic, including domain switches, interrupt mediation, and syscall frame manipulation.
Note that the Syscall Gate can also be implemented as a standalone loader in
the guest kernel to make it compatible with static binaries and more self-contained; we leave it as future work.

\stitle{KVM/QEMU integration.}
To remain compatible with and reuse the existing KVM/QEMU stack, \sys{} introduces some implementation-induced overhead.
First, \sys{} still maintains both the guest page table and the shadow page table.
It enables direct HPA installation (\S\ref{sec:mem}) by hooking guest page-table writes through Linux \texttt{pv\_ops} and adding 22 LoC for hooking page-table reads,
which are needed because hardware-maintained bits (e.g., Accessed and Dirty) are updated only in the shadow page table.
Thus, \sys{} propagates writes to both tables and serves A/D-bits reads from the shadow table by mapping the shadow page table into the guest kernel as read-only.
It also incurs metadata maintenance overheads (evaluated in \S~\ref{sec:eval:syscall_page_fault}),
such as reverse mappings and shadow page table locking,
which a standalone design could avoid.
We leave this optimization to future work.

Second, KVM assumes that a user-space VMM (e.g., QEMU) provides virtual memory regions as guest RAM.
On guest page faults, it invokes \texttt{get\_user\_pages} (GUP) to fault in QEMU-owned pages and install them into the SPT/EPT.
On release, QEMU reclaims pages via \texttt{madvise} with \texttt{MADV\_DONTNEED}, which is costly at fine granularity.
\sys{} allows \pager{} to request host pages directly without mapping them into the QEMU process.
If the host later accesses such memory (e.g., through the host's VirtIO stacks),
the page fault is intercepted and resolved by reusing the existing page if already allocated by \pager{}.
This improves normal memory fault-in and reclamation by avoiding GUP and QEMU page table invalidation, respectively.

\stitle{Limitations.}
One limitation is that MPK-based isolation can be bypassed through control-flow hijacking or misuse of \emph{wrpkru}~\cite{liu2025nanozone},
so prior work enhances it with binary rewriting~\cite{vahldiek2019erim}.
However, binary rewriting can break compatibility and interfere with JIT compilation because of x86 variable-length instructions~\cite{hedayati2019hodor,shi2025hardware}.
\sys{} therefore retains the privileged syscall path as an optional fallback.
Meanwhile, orthogonal work can strengthen intra-address-space isolation, such as using debug registers to monitor sensitive instructions~\cite{hedayati2019hodor,voulimeneas2022you},
combining static binary rewriting with dynamic traps for binary inspection~\cite{vahldiek2019erim}, or reintroducing
control-flow integrity checks to prevent hijacking attacks~\cite{park2019libmpk}. We leave their integration to future work.
Finally, even if the isolation is bypassed, the host, \gates{}, and \pager{} expose only narrow interfaces~(17 calls), which arguably remain sufficient to protect the host.
Another limitation is that some applications may already use MPK for their own purposes~\cite{chen2024limitations}.
In such cases, \sys{} allows application-level configuration during deployment to re-enable address-space-based isolation to trade off isolation overhead against compatibility.

%% file: eval.tex
\section{Evaluation}
\label{sec:eval}

This section presents an evaluation of \sys{} along three axes.
First, we measure system and memory-reclamation performance with microbenchmarks~(\S\ref{sec:eval:microbench}).
Second, we test whether these gains carry over to applications along two traditional cloud workloads: syscall-intensive and memory-intensive workloads~(\S\ref{sec:eval:cloud-app}).
Third, we evaluate both the performance and memory elasticity of agentic workloads~(\S\ref{sec:eval:agent}).

\stitle{Baselines.}
We compare \sys{} with three baselines: \runv{}, PVM~\cite{huang2023pvm}, and \runc{}.
For memory elasticity experiments, we additionally compare against VirtIO-Balloon, VirtIO-Mem, and HyperAlloc~\cite{wrenger2025hyperalloc}.
For PVM and \runv{}, we include huge-page configurations for VM (container) memory.
For \runc{}, we enable side-channel mitigations such as KPTI and IBRS,
since all other VM-based baselines enable these protections at the boundary between containers and the host kernel.

\stitle{Testbed.}
All evaluations are conducted on an AMD server with an EPYC-9654 CPU running at 2.4GHz and 125GB of memory.
For nested-cloud experiments, containers are deployed inside a hardware-assisted L1 VM configured with 32 vCPUs and 64GB of memory.
Both the L0 host and the L1 VM run Ubuntu 22.04 with Linux 6.7.0-rc6.

\subsection{Microbenchmarks}
\label{sec:eval:microbench}

We use microbenchmarks to evaluate \sys{}'s system performance and memory-reclamation performance,
and explain the overheads and optimizations relative to the baselines.

\subsubsection{System Benchmarks}
\label{sec:eval:syscall_page_fault}

\begin{figure}[htb]
    \centering
    \includegraphics[width=1\columnwidth]{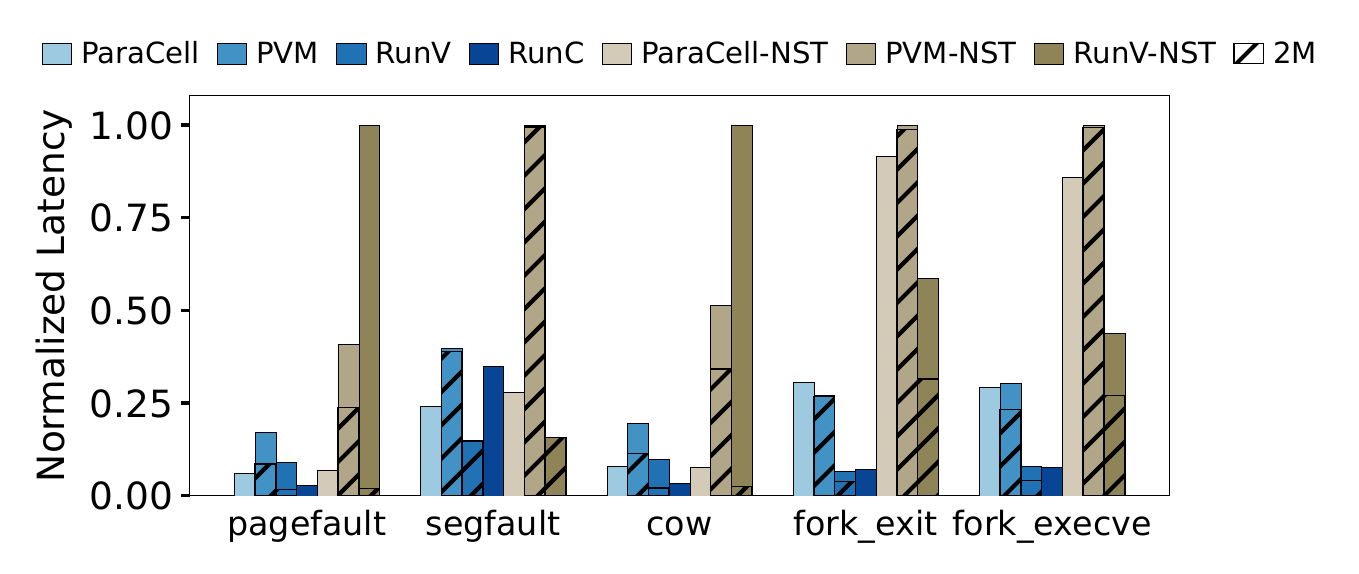}
    \caption{System benchmark latency on LMbench. 2M: 2MB huge-page mapping for VM (container) memory. NST: nested virtualization.}
    \label{fig:eval:syscall_bars_lmbench}
\end{figure}

We first evaluate system performance, including syscalls, page faults, and process management, with LMbench~\cite{lmbench} microbenchmarks.
Table~\ref{tab:eval:syscall_latency} reports raw syscall latency,
and Figure~\ref{fig:eval:syscall_bars_lmbench} summarizes the LMbench results for page faults and process management.

\begin{table}[htb]
    \centering
    \footnotesize
    \caption{Raw syscall~(\texttt{getpid}) latency comparison.}
    \label{tab:eval:syscall_latency}
    \begin{tabular}{cccccc}
        \toprule
        System & \sys{} & \sys{} w/o Depr. & PVM & \runv{} & \runc{} \\
        \midrule
        Latency~(ns) & 107 & 256 & 320 & 96 & 404 \\
        \bottomrule
    \end{tabular}
\end{table}

\stitle{Syscall.}
\sys{} matches \runv{} latency, adding only 11ns for syscall hooking and software emulation of deprivileged syscall instructions~(\S\ref{sec:sgate}).
In contrast, PVM incurs $\sim$210ns overhead from two page-table switches,
while \runc{} adds $\sim$300ns overhead due to side-channel mitigations.
Disabling syscall deprivileging in \sys{}~(\sys{} w/o Depr.) adds $\sim$150ns overhead for the additional privilege-level switch.


\stitle{Page fault.}
On bare metal, \sys{} reduces anonymous page fault latency by 63.9\% compared with PVM, 27.8\% compared with PVM using huge pages, and 31.4\% compared with \runv{} without huge pages.
Under nested virtualization, the reduction grows to 82.8\% compared with PVM and 93.0\% compared with \runv{}.
We next break down the latency of an anonymous page fault.
The total latency of \sys{} is 3991ns.
The additional KVM metadata maintenance accounts for 771ns~(19\%) for the PTE of the faulting user process and 466ns~(12\%) for the direct-mapping PTE,
which could be mitigated as discussed in \S\ref{sec:impl}.
The raw \texttt{set\_pte} accounts for 275ns~(7\%), thanks to the lightweight switch overhead to \pager{}.
The batched GPA$\rightarrow$HPA binding (plus HPA allocation) is amortized to 175ns~(4\%) per page.
Finally, \gates{} reduces the GU/GK switch latency to 1622ns, approaching the \runv{} result of 1028ns.

\sys{} reduces the fault latency of copy-on-write~(CoW) by up to 58.8\% over PVM and 17.1\% over \runv{}, and by 85.0\% and 92.3\% under nested virtualization, respectively.
Segmentation faults do not require a secondary fault for host paging,
so \sys{} improves over PVM by up to 39.6\% on bare metal and 72.2\% under nested virtualization by eliminating page-table switches.
However, it remains slower than \runv{} due to modest page-fault forwarding overhead.
This slowdown is acceptable because segmentation faults are typically infrequent or followed by additional user-level paging,
making the overhead unlikely to dominate end-to-end performance.

\stitle{Process management.}
On bare metal, \sys{} incurs 12.5\%--14.4\% and $-$3.5\%--26.1\% slowdowns over PVM on \texttt{fork}/\texttt{exit} and \texttt{fork}/\texttt{execve}, respectively.
This is because \sys{} delegates page-table writes during page-table cloning to \pager{} through \texttt{pv\_ops},
whereas PVM optimizes this path by allowing direct page-table writes and deferring GPT-to-SPT synchronization to shadow faults.
Compared with \runv{}, \sys{} incurs up to a 3.6\texttimes{} slowdown because its exit path performs KVM metadata maintenance due to a current implementation limitation~(\S\ref{sec:impl}).
We argue that this slowdown is acceptable because \sys{} accelerates the subsequent on-demand paging compared with both PVM and \runv{}.
The trend is similar under nested virtualization, since process management incurs infrequent VM exits.

\subsubsection{Memory Reclamation}

\S\ref{sec:eval:syscall_page_fault} evaluates the cost of faulting pages into a container.
This section evaluates how quickly a system can reclaim container memory to reduce its host footprint.
We measure reclamation throughput over an 8GB region.
In \sys{}, reclamation consists only of guest-kernel page deallocation,
because physical pages are already exposed to the host at the time they are freed.
In the other baselines, reclamation also includes the host-side shrink work,
such as HyperAlloc shrinking, balloon inflation, or hot-unplug,
because these designs rely on a host-initiated operation to discover and remove reclaimable memory.
For HyperAlloc, we include a regular-page setup by disabling huge-page mapping for container memory,
which emulates fine granularity with zero additional metadata overhead.

\begin{figure}[htb]
    \centering
    \includegraphics[width=1\columnwidth]{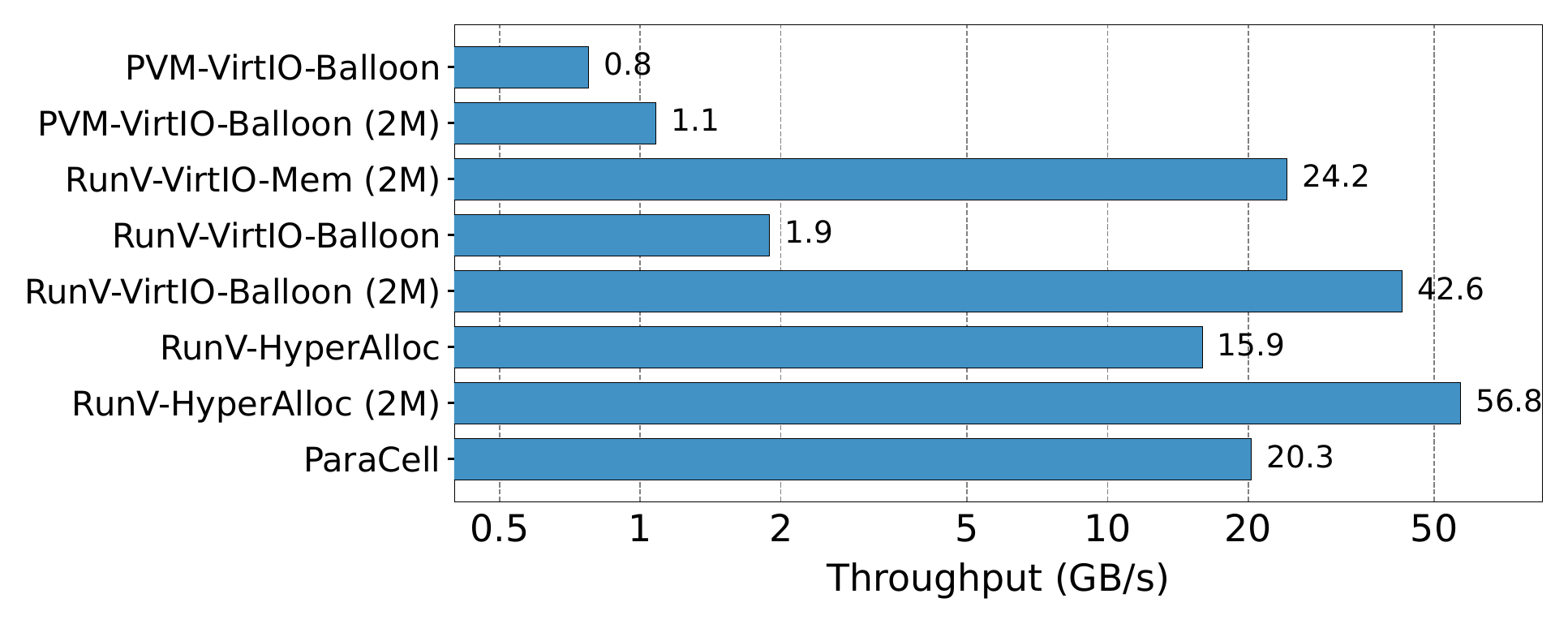}
    \caption{Memory reclamation throughput shown in logarithmic scale. 2M: 2MB huge-page mapping for VM (container) memory.}
    \label{fig:eval:mem_elasticity_micro}
\end{figure}

Figure~\ref{fig:eval:mem_elasticity_micro} shows that \sys{} substantially improves fine-grained reclamation.
Compared with regular-page setups,
\sys{} improves reclamation throughput by $10.7\times$ over \runv{} and $26.2\times$ over PVM with VirtIO-Balloon,
and by $1.3\times$ over regular-page HyperAlloc.
Ballooning pays repeated guest--host transitions and EPT/SPT invalidations at 4KiB granularity.
HyperAlloc avoids the VirtIO path by sharing guest allocator state with QEMU,
but it still performs a host-side scan and per-page invalidation,
whereas \sys{} exposes page deallocation directly on the guest free path.

Compared with huge-page setups, \sys{} trades some raw throughput for fine-grained reclaimability.
Huge-page ballooning and VirtIO-Mem improve reclamation throughput by operating on 2MB chunks,
reaching 42.6GB/s and 24.2GB/s on \runv{}, respectively.
However, VirtIO-Mem is primarily designed for VM expansion;
shrinking can reliably reclaim only memory from the movable zone, often at the cost of memory compaction.
HyperAlloc reaches 56.8GB/s because each operation covers a larger block,
but all these huge-page setups can leave partially used 2MB chunks unreclaimable at the application's allocation granularity (\S\ref{sec:eval:agent}).

\subsection{Traditional Cloud Workloads}
\label{sec:eval:cloud-app}

We next evaluate if the microbenchmark gains translate to traditional cloud workloads.
We organize them into two groups: syscall-intensive and memory-intensive workloads.

\subsubsection{Syscall-Intensive Workloads}

\begin{figure}[htb]
    \centering
    \includegraphics[width=\columnwidth]{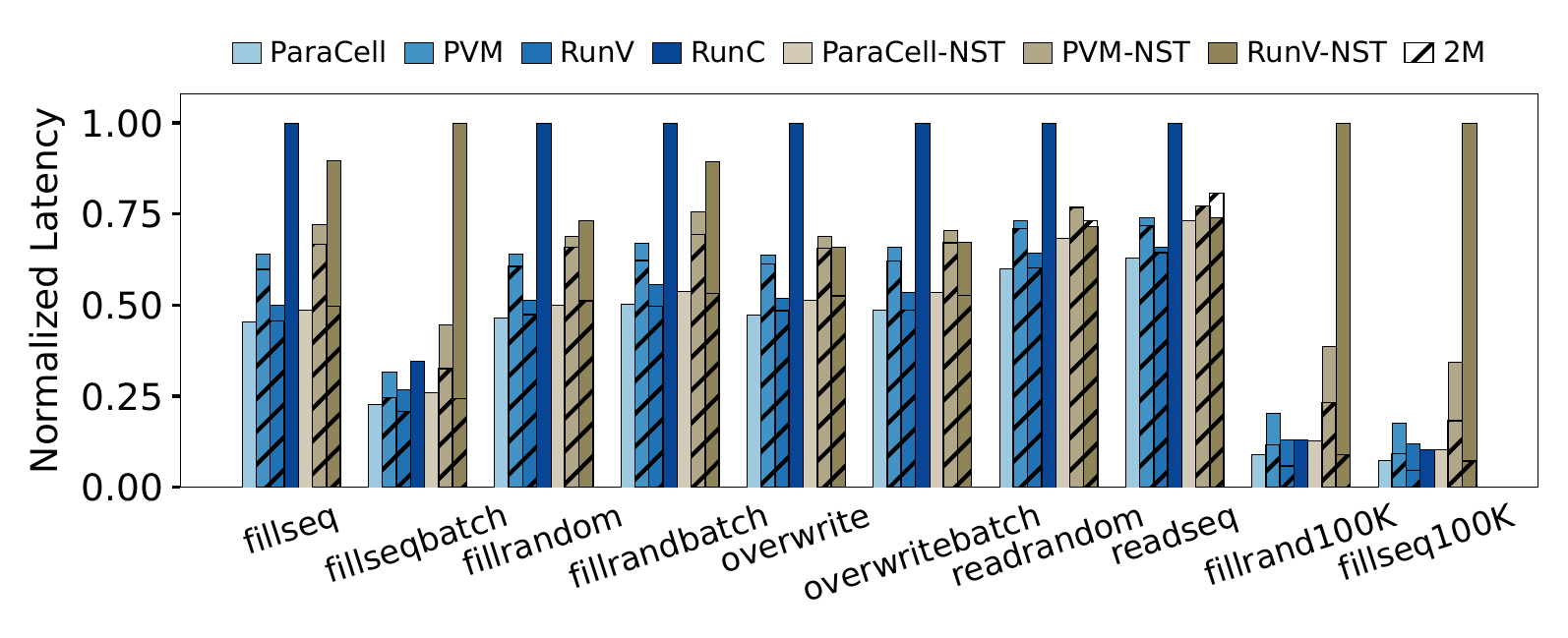}%
    \caption{Normalized latency on sqlite-bench. 2M: 2MB huge-page mapping for VM (container) memory. NST: nested virtualization.}
    \label{fig:eval:sqlitebench}
\end{figure}

We first run sqlite-bench~\cite{sqlite-bench} on SQLite~\cite{sqlite} to measure end-to-end latency dominated primarily by frequent file syscalls.
To isolate storage I/O, we place database artifacts on an in-memory file system.
Figure~\ref{fig:eval:sqlitebench} reports both bare-metal and nested results.

For small fill/overwrite workloads, where syscall overhead dominates, \sys{} reduces latency by up to 28.8\% compared with PVM and matches \runv{} latency,
benefiting from lightweight intra-container isolation. 
For larger batched operations, where page fault overhead becomes more visible, \sys{} reduces latency by up to 33.5\% and 14.9\% compared with PVM and \runv{}, respectively.
On the largest batch~(denoted by the \texttt{100K} suffix), \sys{} reduces latency by up to 58.8\% and 38.9\% over PVM and \runv{}, respectively.
Similar to the slowdown shown in \S\ref{sec:eval:syscall_page_fault},
\runc{} shows significant slowdown even compared with PVM on small and larger-batch workloads, indicating that the overhead of side-channel mitigations in the kernel
can potentially dominate the end-to-end latency of applications if not treated properly.
Nested configurations further amplify the benefit because page faults contribute a larger share of end-to-end time.
\sys{} reduces latency by up to 69.7\% and 89.6\% compared with PVM and \runv{}, respectively.

\subsubsection{Memory-Intensive Workloads}
\label{sec:eval:memory-intensive}

To demonstrate that the microbenchmark improvements of \sys{} generalize to realistic memory-intensive workloads,
we next evaluate applications that are intensive in page faults and TLB misses, drawn from PARSEC~\cite{zhan2017parsec3} and vmitosis~\cite{panwar2021fast}.

\begin{figure}[htb]
    \centering
    \includegraphics[width=\columnwidth]{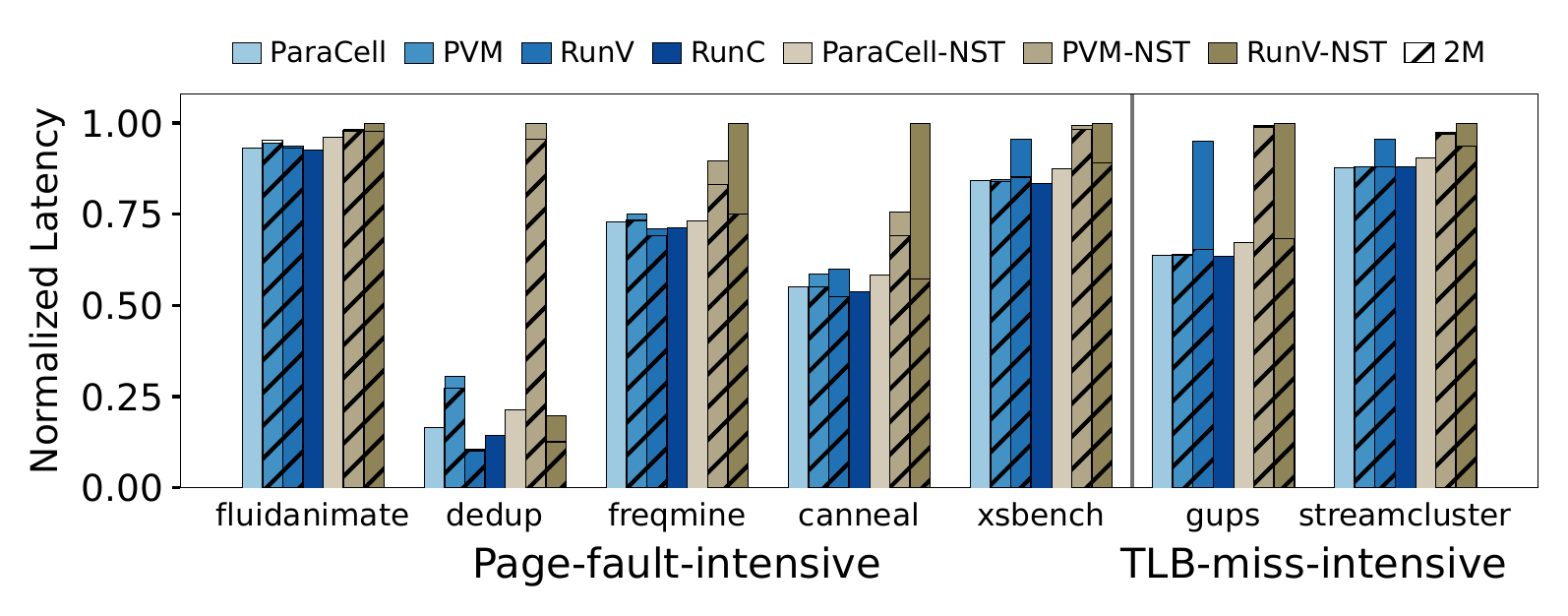}
    \caption{Normalized latency on memory-intensive workloads. 2M: 2MB huge-page mapping for VM (container) memory. NST: nested virtualization.}
    \label{fig:eval:pagefault_apps}
\end{figure}

\stitle{Page-fault-intensive workloads.}
As shown in Figure~\ref{fig:eval:pagefault_apps}, in the bare-metal setting, \sys{} reduces latency by up to 46\% compared with PVM and by up to 39\% compared with PVM (2M).
The larger gap on \texttt{dedup} arises from frequent memory unmaps that trigger TLB flushes.
In PVM, these cause additional TLB-emulation VM exits, whereas \sys{} flushes the TLB directly via \pager{}, avoiding this overhead.
Except for \texttt{dedup}, \sys{} reduces latency by up to 11.9\% compared with \runv{},
while incurring at most 5.4\% and 2.9\% overhead relative to \runv{} (2M) and \runc{}, respectively.
The exception is again \texttt{dedup},
because \runv{} does not return unmapped memory to the host,
and \runc{} does not require an additional kernel direct-mapping TLB flush.
Under nested virtualization, the same trend becomes more pronounced, achieving up to 78.5\% latency reduction.

\stitle{TLB-miss-intensive workloads.}
\sys{}, PVM, and \runc{} show similar performance on the TLB-miss-intensive workloads in both settings.
\runv{} incurs up to 50\% slowdown over the others due to its two-dimensional page walks.

\subsection{Agentic Workloads}
\label{sec:eval:agent}

We evaluate \sys{} on end-to-end agentic workloads using two complementary benchmarks:
SkillsBench~\cite{li2026skillsbench} (applying agent skills) and SWE-bench~\cite{jimenez2023swe} (solving software issues).
SkillsBench provides deterministic ground-truth scripts for each agent skill, enabling accurate end-to-end latency measurements free of agent nondeterminism.
We drive SWE-bench with Claude Code~\cite{anthropic-claude-code} on Sonnet 4.6~\cite{sonnet-4-6} to obtain a genuine agent workload.
We empirically find that memory elasticity is largely insensitive to agent-level nondeterminism.
We therefore use Claude-Code-driven SWE-bench for memory elasticity experiments,
while restricting latency measurements to deterministic SkillsBench.

\begin{figure}[htb]
    \centering
    \includegraphics[width=\columnwidth]{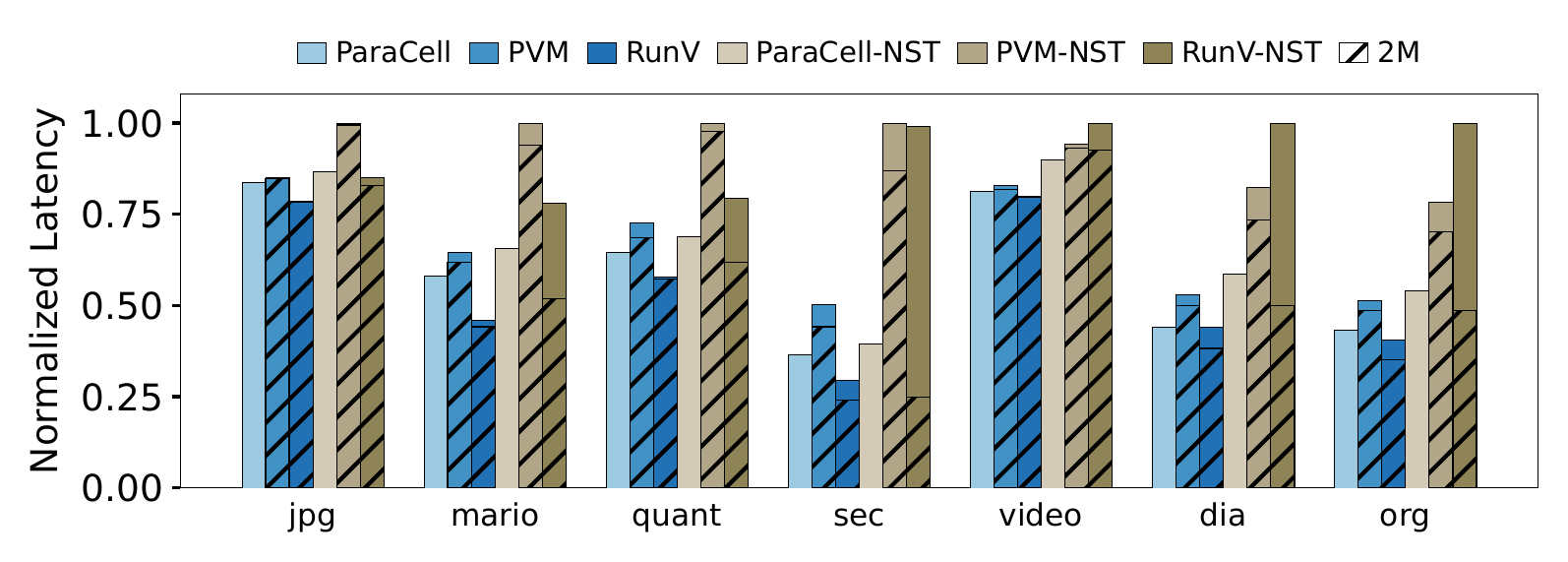}
    \caption{Normalized end-to-end latency on SkillsBench. 2M: 2MB huge-page mapping for VM (container) memory. NST: nested virtualization.}
    \label{fig:eval:skillsbench_latency}
\end{figure}

\stitle{Performance.}
We randomly sample multiple SkillsBench tasks and report normalized latency.
Figure~\ref{fig:eval:skillsbench_latency} shows the latency of tool execution,
while user-perceived completion time also includes LLM inference and agent initialization~\cite{zheng2026agentcgroup}.
Following its study, tool execution accounts for $\sim$40\% of total completion time.
We therefore estimate user-perceived latency by adding the remaining 60\% for LLM inference and agent initialization to the measured tool-execution time,
and report all numbers below in this form.
In the bare-metal setting, \sys{} reduces latency by 5.7\% and 3.4\% on geometric mean, and by up to 15.1\% and 8.3\%, compared with PVM and PVM (2M), respectively.
Compared with \runv{} and \runv{} (2M), \sys{} incurs geometric-mean slowdowns of 3.7\% and 6.5\%, and maximum slowdowns of 8.3\% and 13.8\%, respectively.
The results reflect the worst case for \sys{},
since these baselines never reclaim guest memory,
whereas \sys{} delivers fine-grained reclamation as reported below.
Under nested virtualization, \sys{} outperforms all baselines,
reducing geometric-mean latency by 26.2\%, 21.9\%, 24.8\%, and 1\% over PVM, PVM (2M), \runv{}, and \runv{} (2M), respectively.

\begin{figure}[htb]
    \centering
    \includegraphics[width=\columnwidth]{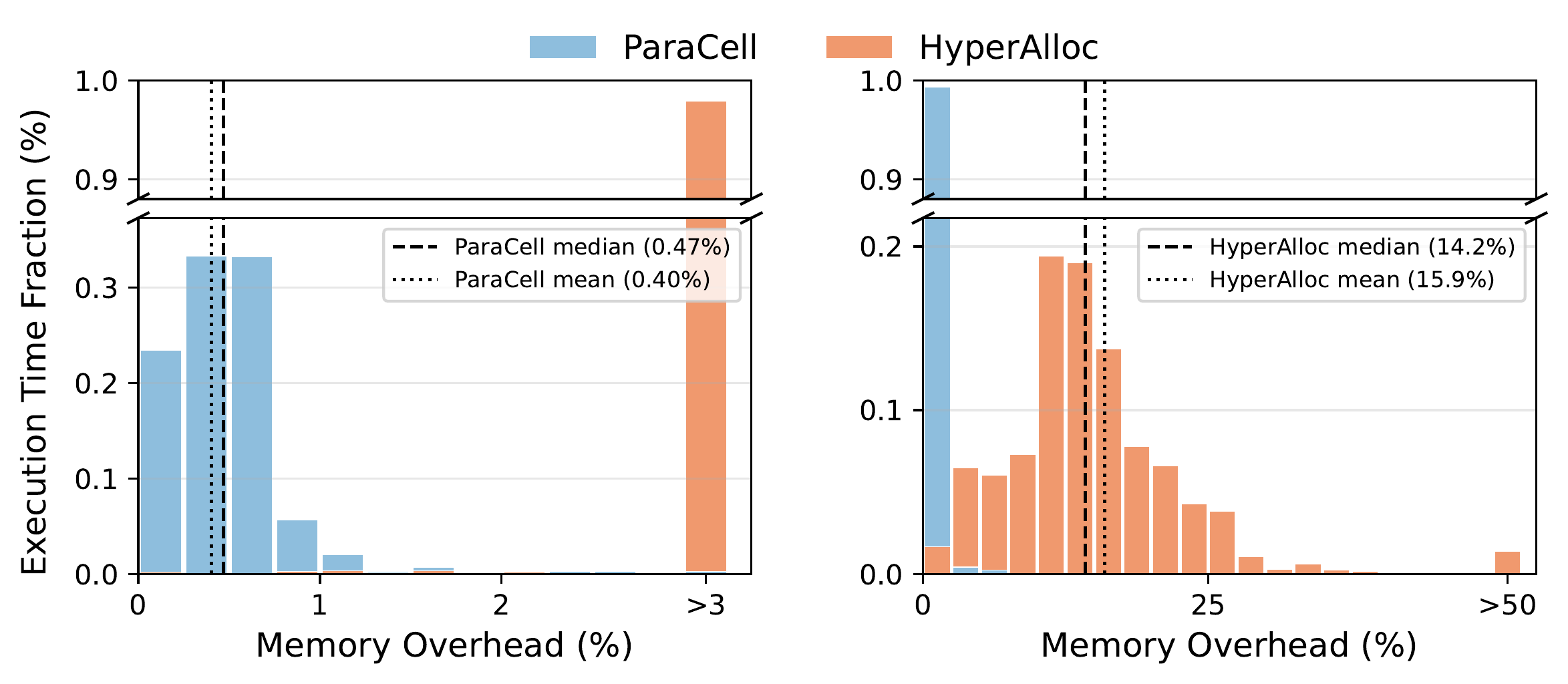}
    \caption{
        Distribution of execution time fraction across host-allocated memory overhead levels during SWE-bench runs,
        shown at two x-axis scales to expose both the low (left) and high (right) overhead regions.
    }
    \label{fig:eval:rss_waste_distribution}
\end{figure}

\stitle{Memory elasticity.}
We sample tasks from SWE-bench multiple times and profile memory at 1Hz during each run to
track both the host-allocated memory and the memory truly consumed by the guest at 4KB granularity.
We define \emph{memory overhead} as the extra host-allocated memory relative to the guest's truly in-use 4KB memory.
On SkillsBench, \sys{}'s host-allocated memory closely tracks guest in-use memory throughout the execution,
resulting in only 0.2\% mean memory overhead.
In contrast, HyperAlloc suffers from internal fragmentation within huge pages,
leading to 35.8\% mean memory overhead.
Similar trends can be observed in SWE-bench results.
Figure~\ref{fig:eval:rss_waste_distribution} shows the distribution of normalized execution time across different memory overhead levels.
\sys{} achieves a mean memory overhead of only 0.4\%,
with nearly all execution time concentrated below 1\% memory overhead.
In contrast, HyperAlloc incurs a mean overhead of 15.9\% and spends a noticeable execution time fraction (2\%) above 50\% memory overhead.

%% file: related-work.tex
\section{Other Related Work}
\label{sec:related-work}

\stitle{Intra-address-space isolation.} 
Intra-address-space isolation is a technique to achieve lightweight isolation across different components within a single address space.
Various systems leverage existing hardware protection primitives to implement such isolation for different purposes.
ERIM~\cite{vahldiek2019erim} and Hodor~\cite{hedayati2019hodor} focus on protecting userspace data-plane libraries from
calling applications and each other without requiring device-level resource isolation.
BULKHEAD~\cite{guo2024bulkhead} targets kernel compartmentalization, isolating untrusted kernel components from each other using PKS.
LightEnclave~\cite{gu2022hardware}
extends this idea to the enclave level and proposes a hardware--software co-design to reduce the TCB size for each lightweight TEE enclave.
UnderBridge~\cite{gu2020harmonizing} harmonizes microkernel services within a single address space and 
reduces the IPC switch overhead between each microkernel service using PKS.
CKI~\cite{shi2025hardware} extends the hardware feature to create an isolated intra-kernel domain for guest and host kernels
without relying on hardware virtualization, optimizing the performance of nested virtualization.

In contrast, \sys{} reintroduces intra-address-space isolation into paravirtualized containers to reduce the domain switch overhead between
the guest kernel and the guest user while still preserving the isolation invariants.

\stitle{VM memory elasticity.}
Prior VM memory-elasticity mechanisms differ mainly in how they bridge the guest--host semantic gap around reclaimable pages.
Classic memory ballooning~\cite{waldspurger2002memory,schopp2006resizing} reclaims memory by inflating a guest-side balloon driver and letting the guest OS choose which pages to surrender.
This preserves guest policy, but reclamation still depends on explicit host--guest coordination and can react poorly to bursty demand.
VirtIO-Mem~\cite{hildenbrand2021virtio} resizes VMs through memory hotplug and hot-unplug,
but it works at coarse block granularity.
Hyperupcalls~\cite{amit2018design} reduces coordination overhead by letting the hypervisor execute guest-supplied logic to inspect allocator state directly,
while V-Probe~\cite{wang2023efficient} and HyperAlloc~\cite{wrenger2025hyperalloc} push this direction further through active page-metadata inspection and allocator co-design.
Orthogonally, transcendent memory~\cite{magenheimer2009transcendent} uses host memory as auxiliary swap or cache space and requires explicit adaptation to the guest programming model.

In contrast, \sys{} places elasticity directly on the guest free path and targets page-granular reclaim with low boundary overhead,
without relying on balloon inflation, coarse hot-unplug units, or wholesale allocator replacement.

\stitle{Nested virtualization optimizations.}
Nested virtualization has long suffered from excessive world switches between the bare-metal hypervisor~(L0),
guest hypervisor~(L1), and nested guest~(L2)~\cite{ben2010turtles},
and prior work attacks this cost along four complementary directions.
Some work~\cite{lim2017neve,vilanova2019using} relies on specialized hardware to reduce the frequency or cost of VM exits,
but these features are not widely available.
DVH~\cite{lim2020optimizing} builds an I/O fast path that exposes L0 virtual devices directly to L2,
removing L1 from the I/O critical path at the cost of surrendering L1's policy enforcement on L2 I/O.
HyperTurtle~\cite{zur2025accelerating} employs code offloading,
executing selected L1 logic as verified eBPF programs inside L0 to reduce world switches,
but is constrained by eBPF expressiveness.
Other work~\cite{ye2023free,williams2016enabling,bitchebe2022out} sidesteps nesting altogether by eliminating or repurposing L1,
at the cost of reduced hypervisor-level flexibility and orchestration.

\sys{} adopts a paravirtualization design and uses MPK to accelerate guest privilege transitions,
avoiding the costs caused by missing hardware virtualization in nested clouds.

\stitle{Container optimizations for specific workloads.}
A large body of work optimizes the container stack for different workload classes.
For serverless functions, where cold-start latency and deployment density dominate,
prior work slims down the container and microVM~\cite{oakes2018sock,li2022rund,agache2020firecracker} stacks,
or accelerates instantiation via checkpoint restore~\cite{du2020catalyzer,alquraan2026drops}, RDMA-based remote fork~\cite{wei2023mitosis},
and cross-function environment sharing~\cite{huang2024trenv,segarra2026nanvix}.
For microservices with microsecond-scale tail-latency targets,
prior work reduces per-invocation overheads through microsecond-scale scheduling~\cite{jia2021nightcore},
eBPF-based shared-memory dataplanes~\cite{qi2022spright}, and kernel-bypass I/O~\cite{fried2024making}.
For LLM agents, AgentCgroup~\cite{zheng2026agentcgroup} characterizes and controls OS-level resource usage of agent processes,
exposing gaps in existing cgroup-based isolation.

\sys{} improves general container mechanisms,
particularly performance-critical boundary crossings and memory elasticity,
in a workload-agnostic manner,
creating opportunities to incorporate existing optimizations.

%% file: conclusion.tex
\section{Conclusion}
\label{sec:conclusion}
\sys{} is a paravirtualized secure container runtime.
\sys{} enables lightweight intra-container isolation with MPK and fine-grained memory elasticity by exposing guest-kernel memory-management intent.
Evaluation shows promising performance and memory utilization across traditional cloud workloads and emerging agent workloads.
\sys{} points to a broader design direction in which the host cooperates with guest-kernel abstractions, rather than inferring them indirectly, to build secure runtimes that are both faster and more memory-efficient.

%% file: appendix.tex
\appendix